\documentstyle[12pt]{article}

\begin{document}
\title{Absorption in a Particle Oscillations}
\author{V.A.Kuzmin, V.I.Nazaruk\\
Institute for Nuclear Research of RAS, 60th October\\
Anniversary Prospect 7a, 117312 Moscow, Russia.*}

\date{}
\maketitle
\bigskip

\begin{abstract}

In the framework of field approach with the finite time interval the particle
oscillations in a medium are considered. The absorption in a final states do
not lead to dramatic suppression of $ab$ transitions. Also we touch on the
problem of infrared singularities. The approach under study is infrared-free.
It is shown that "correction" published recently [A.Gal, Phys. Rev. {\bf C61} 
(2000) 028201] is clearly wrong.
\end{abstract}

\vspace{12cm}
*E-mail addresses: kuzmin@ms2.inr.ac.ru; nazaruk@al20.inr.troitsk.ru

\newpage
\setcounter{equation}{0}
\section{Introduction}
The theory of $ab$ oscillations [1] are based on single-particle model. The
interaction of particles $a$ and $b$ with the matter is described by potentials
$U_{a,b}$. $ImU_b$ is responsible for loss of $b$-particle intensity. The wave
functions $\Psi _{a,b}$ are given by equations of motion. The index of
refraction, the forward scattering amplitude $f(0)$ and potential are related
to each other, so later on the standard approach is referred to as potential
model.

In some instances there is a need to consider the $ab$ conversion in the
matter followed by reaction
\begin{equation}
(a-\mbox{medium})\rightarrow (b-\mbox{medium})\rightarrow b+c\rightarrow f.
\end{equation}
Here $c$ is the particle of medium (one should sum over all $c$-particles);
$b+c\rightarrow f$ represents the reaction. The whole process ($ab$ transition,
$b$-medium interaction) takes place in the same layer of matter. $b$-particle
absorption is essential and $ImU_b$ cannot be ignored. An example is the
$n\bar{n}$ transitions [2] in the medium followed by annihilation
\begin{equation}
(n-\mbox{medium})\rightarrow (\bar{n}-\mbox{medium})\rightarrow f,
\end{equation}
where $f$ are the annihilation products which should be detected (see Fig.1a).
Standard calculation (see, for example [3,4] and Section 2) predicts a 
dramatic suppression of this process due to $\bar{n}$-medium interaction. 
Below it is shown that process (2) is qualitatively equivalent to free-space 
process given in Fig.2:
\begin{equation}
n\rightarrow \bar{n}\rightarrow \bar{p}e^+\nu.
\end{equation}
Therefore, process (3) should be drastically suppressed by the decay in a 
final state, what is incorrect. We cite a body of other arguments (see Sec.6)
which point to the fact that potential approach is reasonable as starting 
point only.

In the framework of field approach with finite time interval (FTA) we perform
the direct calculation of the processes shown in Fig.1. The connection
between FTA and $S$-matrix theory is studied. The approach reproduces all the
results in neutrino oscillations in which $Imf_b(0)$ is ignored as well as
standard calculation of the process (2). However, we disagree with standard
calculation, because, in our opinion, the absorption is described improperly in
this case.

\section{Potential model}
In the standard approach the $n\bar{n}$ transitions in the medium are
described by Schrodinger equations
\begin{eqnarray}
(i\partial_t-H_0)\Psi _n(x)=\epsilon \Psi _{\bar{n}}(x),\nonumber\\
(i\partial_t-H_0-V)\Psi _{\bar{n}}(x)=\epsilon \Psi _n(x),\nonumber\\
H_0=-\nabla^2/2m+U_n,\nonumber\\
V=U_{\bar{n}}-U_n=ReU_{\bar{n}}-i\Gamma /2-U_n,
\end{eqnarray}
$\epsilon =(m_2-m_1)/2=1/\tau _{n\bar{n}}$. Here $m$ and $U_n=const$ are the
neutron mass and potential, respectively; $U_{\bar{n}}$ and $\Gamma \sim 100$ 
MeV are the optical potential and annihilation width of $\bar{n}$; $\tau _{n
\bar{n}}$ is a free-space $n\bar{n}$ oscillation time; $m_{1,2}$ are the 
masses of the stationary states $n_{1,2}$. $\epsilon $ is a small parameter. 
The initial conditions are $\Psi _n(0,{\bf x})=n(0,{\bf x})$, $\Psi _{\bar{n}}
(0)=0$, where $n(x)=V^{-1/2}\exp (-i\epsilon _nt+i{\bf p}_n{\bf x})$,  
$\epsilon _n ={\bf p}_n^2/2m+U_n$.

For analogy with field approach we introduce the evolution operator
$U(t)=I+iT(t)$. For $V=const.$ in the lowest order in $\epsilon$ we obtain 
matrix elements $U_{ii}(t)=<\!n(0)\!\mid \!\Psi_n(t)\!>$ and $T_{\bar{n}i}(t)=
<\!n(0)\!\mid \!\Psi _{\bar{n}}(t)\!>$:
\begin{eqnarray}
T_{\bar{n}i}(t)=(\epsilon /V)(exp (-iVt)-1),\nonumber\\
T_{ii}(t)=i(\epsilon/V)^2[1-iVt-\exp (-iVt)].
\end{eqnarray}
The probability to find antineutron $W_{\bar{n}}(t)$ is
\begin{equation}
W_{\bar{n}}(t)=\mid\!T_{\bar{n}i}(t)\!\mid ^2=(\epsilon /\mid\! V\!
\mid )^2[1-2\cos(ReVt)e^{-\Gamma t/2}+e^{-\Gamma t}].
\end{equation}
The unitarity condition gives
\begin{equation}
1=\mid \!U_{ii}(t)\!\mid ^2+W_{\bar{n}}(t)+W_{ann}(t),
\end{equation}
where $W_{ann}(t)$ is the probability to find the annihilation products. In the
potential model the probability of $n\bar{n}$ conversion $W_{pot}$ is defined
by equation
\begin{equation}
W_{pot}(t)=1-\mid \!U_{ii}(t)\!\mid ^2=2ImT_{ii}(t).
\end{equation}
For brevity, we put below
\begin{equation}
V=-i\Gamma /2,\;\;\; \Gamma t\gg 1.
\end{equation}
Then
\begin{equation}
W_{pot}(t)\approx 4\epsilon ^2t/\Gamma .
\end{equation}
Noting that free-space $n\bar{n}$ transition probability  is
\begin{equation}
W_f=\epsilon^2t^2
\end{equation}
for suppression factor $R_{pot}$ we have $R_{pot}=W_{pot}/W_f=4/\Gamma t\ll 1$.

We would like to stress that Eq.(7) is automatically fulfilled only for correct
model with Hermitian operators. Also recall that instead of realistic operator
of $\bar{n}$-medium interaction $\hat{U}_{\bar{n}}$ the parametrization
\begin{equation}
\hat{U}_{\bar{n}}=ReU_{\bar{n}}-i\Gamma /2=\mbox{const }
\end{equation}
is used. On the reasons given below we abandon the approximation (12). In this
case the usual procedure connected with diagonalization of mass matrix cannot
be realized. However, the result do not depend on the basis. All existing
calculations have been done in $n-\bar{n}$ representation.

\section{Field approach}
Let us consider the process (2) in the framework of field approach. The 
interaction Hamiltonian involves two terms:
\begin{eqnarray}
H_{n\bar{n}}(t)=\epsilon \int d^3x(\bar{\Psi }_{\bar{n}}\Psi _n+H.c.),
\nonumber\\
H(t)=(\mbox{all }\bar{n}-\mbox{medium interactions}) - U_n,
\end{eqnarray}
$H_I=H_{n\bar{n}}+H$. Here $\Psi _n$ and $\Psi _{\bar{n}}$ are the fields of
$n$ and $\bar{n}$, respectively; $m_{\bar{n}}=m_n=m$. The background nuclear
matter field $U_n$ is included in unperturbed Hamiltonian $H_0$; quadratic
term $H_{n\bar{n}}$ included in $H_I$. The sole physical distinction with
model (4) lies in description of $\bar{n}$-medium interaction $H$. If it is
putted that $H(t)=\int d^3x(-i\Gamma /2)\bar{\Psi }_{\bar{n}}\Psi _{\bar{n}}$
(what is at least unjustified), then the potential model results are
reproduced (see below). Field approach allows one to calculate directly the
off-diagonal terms.

Let us $U_n=ReU_{\bar{n}}=0$, $\Gamma =\Gamma _{\beta }$, $H=H_{\beta} $, 
where $\Gamma _{\beta }$ and $H_{\beta }$ are the width and Hamiltonian of 
free-space decay $\bar{n}\rightarrow \bar{p}e^+\nu $. In this case Eqs.(4) and 
(13) describe the free-space process (3) in the framework of single-particle 
model and field approach, respectively. (As with process (2), the neutron 
decay is excluded. In this connection the free-space process (3) is imaginary 
one, which is inessential.) Then according to Eq.(10) the process (3) is 
drastically suppressed: $W_{pot}(t)\approx 4\epsilon ^2t/\Gamma _{\beta }$. So 
in the field approach this process (see Fig.2) should be also suppressed by 
the decay in a final state, what seems unrealistic.

For Fig.1a (and, therefore, for Fig.2) it was shown [5] that the process
probability is $W(t)\approx W_f(t)$. Below this result is derived from
consideration of more general process shown in Fig.1b. However, first of all
we verify FTA, namely, reproduce the potential model results (5).

\section{Calculation}
First of all we consider the $n\bar{n}$ transitions with $\bar{n}$ in the
final states ($\bar{n}$ are detected). The similar problem takes place in
neutrino oscillations. Due to the zero momentum transfer in the vertex
corresponding to $H_{n\bar{n}}(t)$ the process amplitude is singular
$$
M_s=\epsilon \frac{1}{\epsilon_n-{\bf p}^2_n/2m-U_n}M\sim 1/0,
$$
where $M$ is the amplitude of $\bar{n}$-medium interaction. (The amplitude of
potential model is considered in Sec.5.) For solving the problem the approach 
with finite time interval was proposed [4]. In the lowest order in $\epsilon $ 
we have
$$
<\!\bar{n}0\!\mid U(t)-I\mid\!0n\!>=iT_{\bar{n}i}(t)=(-i)
<\!\bar{n}_p0\!\mid \int_{0}^{t}dt_{\beta }H_{n\bar{n}}(t_{\beta })+
T^{\bar{n}}(t-0)\int_{0}^{t_k}dt_{\beta }H_{n\bar{n}}(t_{\beta })\mid\! 0n_p
\!>,
$$
$$
T^{\bar{n}}(t-t_{\beta })=\sum_{k=1}^{\infty}(-i)^k \int_{t_
{\beta }}^{t}dt_1...\int_{t_{\beta }}^{t_{k-1}}dt_kH(t_1)...H(t_k),
$$
where $\mid \!0n_p\!>$ and $\mid\!0\bar{n}_p\!>$ are the states of the medium
containing the neutron and antineutron with 4-momenta $p=(\epsilon _n,{\bf p}_
n)$. Taking into account that $H_{n\bar{n}}\mid\!0n_p\!>=\epsilon
\mid\!0\bar{n}_p\!>$, we change the order of integration [4] and obtain
\begin{eqnarray}
T_{\bar{n}i}(t)=-\epsilon t -\epsilon \int_{0}^{t}dt_{\beta }
iT^{\bar{n}}_{ii}(t-t_{\beta }),\\
iT^{\bar{n}}_{ii}(\tau)=<\!\bar{n}_p0\!\mid T^{\bar{n}}(\tau) \mid\! 0
\bar{n}_p\!>,\nonumber
\end{eqnarray}
where $\tau =t-t_{\beta }$. The $\bar{n}$-medium interaction is separeted out
in block $T^{\bar{n}}_{ii}(\tau)$.

For verification of FTA we calculate $T^{\bar{n}}_{ii}$ in the framework of 
potential model: $V=const $, $H(t)=V(t)=\exp (iH_0t)V\exp
(-iH_0t)=V$. We have
\begin{equation}
iT^{\bar{n}}_{ii}(\tau)=U^{\bar{n}}_{ii}(\tau)-1=\exp (-iV\tau)-1.
\end{equation}
Substituting this expression in Eq.(14) one obtains Eq.(6). This means that
FTA should reproduce all the results in neutrino oscillations. (Commonly, in
neutrino oscillations $Imf_b(0)$ is ignored.) One further important test of FTA
(calculation of $T_{ii}(t)$) have been given in [5]. Therefore, the FTA was
{\em verified} by the example of exactly solvable potential model. It is 
involved in $<\!\bar{n}_p0\!\mid T^{\bar{n}}(\tau) \mid\! 0\bar{n}_p\!>$ as a
special case.

Let us consider the process (2) wherein annihilation products are detected,
namely, the $n\bar{n}$ transitions in the nuclear matter (Fig.1a). Here,
$\Gamma \sim 100$ MeV and $\bar{n}$ inevitably annihilates. We consider the
more general problem. We calculate Fig.1b on the interval $(t/2,-t/2)$. As a
result it will be shown that: (a) When $q\rightarrow 0$ ($q$ is 4-momenta of
particle escaped in the $n\bar{n}$ transition vertex), the result converts to
one corresponding to Fig.1a, which is interesting for us. (b) When $q\neq 0$
and $t\rightarrow \infty $, the result coincides with $S$-matrix one. Such
scheme allows to verify and study the FTA. (c) The functional structure of 
$W_{pot}(t)$ is wrong.

Let us consider the imaginary process
\begin{equation}
n\rightarrow \bar{n}+\Phi.
\end{equation}
For decay to be permissible in vacuum put $m_{\bar{n}}=m-2m_{\Phi }$. The
corresponding process in nuclear matter is shown in Fig.1b. This is a nearest
analogy to the process under study. Instead of Eqs.(13) we have $H_I=H'_{n\bar
{n}}+H$,
\begin{equation}
H'_{n\bar{n}}(t)=\epsilon '\int d^3x(\bar{\Psi }_{\bar{n}}\Phi ^* \Psi _n+
H.c.),
\end{equation}
where $H'_{n\bar{n}}$ is the Hamiltonian corresponding to decay (16). (For
Fig.1a $H'_{n\bar{n}}\rightarrow H_{n\bar{n}}$.) In a manner like the
calculation of $T_{\bar{n}i}(t)$ we have
\begin{eqnarray}
T_b(t)=-<\Phi _qf\mid T^{\bar{n}}(t)\int_{-t/2}^{t_k}dt_{\beta }H'_{n\bar{n}}
(t_{\beta })e^{-\alpha \mid\!t_{\beta }\!\mid } \mid\!0n_p\!>,\\
T^{\bar{n}}(t)=\sum_{k=1}^{\infty}(-i)^k \int_{-t/2}^{t/2}dt_1...\int_{-t/2}
^{t_{k-1}}dt_kH(t_1)...H(t_k)=T\exp (-i\int_{-t/2}^{t/2}dt_1H(t_1))-1.
\end{eqnarray}
Here  $<\!f\!\mid$ represents the annihilation products with $(n)$ mesons.
Multipliere $\exp(-\alpha \mid\!t_{\beta }\!\mid )$, $\alpha >0$ is introduced
for realization of adiabatic hypothesis. One obtains
\begin{eqnarray}
T_b(t)=i\epsilon '\frac{1}{\Delta q-i\alpha }<f\mid T^{\bar{n}}(t)\mid\!0\bar
{n}_{p-q}\!>NF,\\
F=e^{-\alpha \mid t_k\mid }-e^{-\alpha t/2}e^{-i\Delta q(t_k+t/2)},
\end{eqnarray}
$\Delta q=q_0-2m_{\Phi }+({\bf p}_n-{\bf q})^2/2m_{\bar{n}}-{\bf p}^2/2m$. 
Here $\mid\!0\bar{n}_{p-q}\!>$ is the state of the medium containing the 
$\bar{n}$ with 4-momenta $p-q$. The normalizing factors of the wave functions 
of $\bar{n}$ and annihilation mesons are included in $<f\mid T^{\bar{n}}(t)
\mid \!0\bar{n}_{p-q}\!>$ and the other those in the multiplier $N$.

{\em Nonsingular diagram.} - Since $\Delta q\neq 0$ the limit $t\rightarrow
\infty $ can be considered. Then $F=\exp(-\alpha \mid\!t_k\!\mid)$ and $T_b$ is
the usual $S$-matrix element. FTA {\em reproduces} the $S$-matrix result,
{\em as we set out to prove.}

It is easy to estimate the widths corresponding to Fig.1b and free-space decay
(16):
\begin{eqnarray}
\Gamma_b\approx \epsilon'^2\Gamma /(2\pi ^2),\nonumber\\
\Gamma_{free}\approx \epsilon'^2m_{\Phi }/(2\pi ),\nonumber
\end{eqnarray}
where we have put $m_{\Phi }/m\ll 1$. The $t$-dependence is determined by 
exponential decay law
$$
W_{b,free}=1-e^{-\Gamma _{b,free}t}\sim \Gamma _{b,free}t.
$$
These formulas will be needed below.

{\em Singular diagram.} - Let us return to $T_b(t)$ and consider the formal
limit $q\rightarrow 0$. For $n\bar{n}$ transition Hamiltonian the adiabatic
hypothesis should not be used: $\alpha =0$ (see below). Now factor $F/\Delta q
$ is
$$
[1-(1-i\Delta q(t/2+t_k))]/\Delta q=i\int_{-t/2}^{t_k}dt_{\beta }.
$$
Changing the integration order [4] and going to interval $(t,0)$ one obtains
\begin{equation}
T_b(t)=-\epsilon 'N\int_{0}^{t}dt_{\beta }<\!f\!\mid T^{\bar{n}}(t-t_{\beta })
\mid\! 0\bar{n}_p\!>.
\end{equation}
At a point $q=0$, $H'_{n\bar{n}}(t)=H_{n\bar{n}}(t)$ and $N=1$. We have
\begin{eqnarray}
T_b(t)\rightarrow T_a(t)=-\epsilon \int_{0}^{t}d\tau iT^{\bar{n}}_{fi}
(\tau ),\nonumber\\
iT^{\bar{n}}_{fi}(\tau )=<\!f\!\mid T^{\bar{n}}(\tau )\mid\! 0\bar{n}_p\!>.
\end{eqnarray}
$T^{\bar{n}}_{fi}(\tau )$ is an exact annihilation amplitude. $T_a(t)$
coincides with the second term of Eq.(14) with the replacement $<\!i\!\mid =
<\!\bar{n}_p0\!\mid \rightarrow <\!f\!\mid $. This can be considered as a test
for $T_a(t)$. (In the Ref.[5] only Fig.1a was calculated. Also it was shown
that there is a double counting in the potential model.)

\section{Infrared singularities and S-matrix problem formulation}
The amplitude of potential model obtained by means of $S$-matrix approach is
not singular. From microscopic theory standpoint the reason is as follows. Due
to zero momentum transfer in the $\epsilon $-vertex this amplitude contains
singular propagator of $\bar{n}$. However, it also contains block $T^{\bar{n}}_
{ii}$ which is a sum of zero angle rescattering diagrams of $\bar{n}$. As a 
result the self-energy part $\Sigma=V$ appears (see Eq.(22) of Ref.[5]). We 
are interesting in off-diagonal matrix elements which do not contain 
above mentioned sum ($T^{\bar{n}}_{fi}$ instead of $T^{\bar{n}}_{ii}$) and
hence diverges, because one singular propagator after $\epsilon $-vertex
appears in any case. (Note, that formal sum of series  over $\epsilon $ gives
meaningless self-energy part $\Sigma \sim 1/0$.)

FTA is infrared-free. It naturally connected with conditions of experiment.
Really, measurement of any process corresponds to some interval $\tau $. So it 
is necessary to calculate $U_{fi}(\tau )$. The replacement $U(\tau )
\rightarrow S(\infty )$ is justified when the main contribution gives some 
region $\Delta <\tau $, so that
\begin{equation}
U_{fi}(\tau >\Delta )=U_{fi}(\infty )=U_{fi}=S_{fi}=const.
\end{equation}
The expressions of this type are the basis for all $S$-matrix calculations.
In principle, the three cases are possible.

1. There is bound to be asymptotic regime, however it is not achieved
automatically. Then the adiabatic hypothesis is used and usual scheme realizing
in the field theory or non-stationary theory of scattering takes place. Fig.1b
corresponds to this case.

2. There is no asymptotic regime. Example is provided by oscillation
Hamiltonian $H_{n\bar{n}}(t)$. $S$-matrix approach is inapplicable. We have
usual non-stationary problem. Because of this, for Fig.1a the result has been
obtained in the framework of FTA. In the vertex corresponding to $H_{n\bar{n}}
(t)$ the adiabatic hypothesis was not used.

3. For the interaction Hamiltonian $H_I(t)$ the asymptotic is reached
automatically without resort to adiabatic hypothesis. An example of this type
is $T^{\bar{n}}_{ii}(\tau)$: $T^{\bar{n}}_{ii}(\tau \gg 1/\Gamma)=i$.

For non-diagonal matrix elements the asymptotic should be reached as well:
\begin{equation}
T^{\bar{n}}_{fi}(\tau >1/\Gamma )=T^{\bar{n}}_{fi}=const.
\end{equation}
Really, annihilation of $\bar{n}$ in nuclear matter can be considered as a 
system decay with the characteristic time $\sim 10^{-23}$ s. On the other hand 
the observation time $t=T_0\sim 1$ yr [6] is far more then that in the 
experiments on particle decays. So in our case expression (24) is more obvious 
then for any free-space reaction or decay. Recall that all calculations in the 
particle physics based on condition (24). (For process (3) it has a form $T^
{\beta }_{fi}=const$, where $T^{\beta }$ is matrix of $\beta $-decay.) Besides,
for $T^{\bar{n}}_{ii}$ it was verified by direct calculation.

Since $t\gg 1/\Gamma $, by means of (25) one obtains
\begin{equation}
T_a(t)\approx -i\epsilon tT_{fi}^{\bar{n}}.
\end{equation}
$T_{fi}^{\bar{n}}$ can be calculated in the framework of $S$-matrix theory.
However, in our case we know that the $\bar{n}$-nucleus decay probability is
$W^{\bar{n}}=\sum_{f\neq i}\mid T_{fi}^{\bar{n}}\mid ^2=1$. Finally
\begin{equation}
W_{ann}(t)\approx \sum_{f\neq i}\mid -i\epsilon tT_{fi}^{\bar{n}}\mid ^2
=\epsilon^2t^2W^{\bar{n}}=\epsilon^2t^2=W_f.
\end{equation}
Due to the annihilation channel $n\bar{n}$ conversion is practically 
unaffected by the medium. So $\tau_{n\bar{n}}\sim T_{n\bar{n}}$, where $T_{n
\bar{n}}$ is the oscillation time of neutron bound in a nucleus. The 
t-dependence of $W_{ann}$ is determined by that of more slow subprocess of the 
$n\bar{n}$ conversion $W_f\sim t^2$. Formally, quadratic dependence follows 
from expression for $T_a(t)$.

All the results have been obtained by means of formal expansions. They are
valid at any finite $t$. Consequently, the singularity of $S$-matrix amplitude
$M_s$ is a direct result of incorrectness of the formulation of the $S$-matrix
problem. If $t\rightarrow \infty $ Eq.(26) diverges just as $M_s$ does.
So infrared singularities point to the fact that there is no asymptotic
regime. Conversely, the analysis of condition (24) suggests the method for
handling. The lack of asymptotic leads to FTA which is infrared-free. (There
is no asymptotic regime for free-space $K^0\bar{K}^0$-oscillations as well. In
our opinion, it makes sense to look at the calculation of $\Delta m=m_L-m_S$
(GIM mechanism) from this standpoint.)

\section{Discussion}
From Eq.(10) it is seen that:

(1) $W_{pot}\sim 1/\Gamma$, whereas Fig.1 gives the inverse dependence
$\mid T_{a,b}\mid ^2\sim  \mid T_{fi}^{\bar{n}}\mid ^2$ (see Eqs.(20),\\
(26)). Such structure is typical and for $f\neq i$ uniquely determined.

(2) The $t$-dependence of the process probability in the medium and vacuum is
different: $W_{pot}\sim t$, $W_f\sim t^2$. It is beyond reason to such
fundamental change. In our calculation the $t$-dependence is the same: when
$q=0$ (Fig.1a), $W_{ann}\sim t^2$ and $W_f\sim t^2$; when $q\neq 0$ (Fig.1b),
$W_b\sim t$ and $W_{free}\sim t$.

(3) In the medium the $n\bar{n}$ transition suppressed by a factor $R_{pot}=
W_{pot}/W_f=4/\Gamma t$. If $t=T_0=1.3$ yr [6] ($T_0$ is observation time in
proton-decay type experiment) and $\Gamma =100$ MeV then $R_{pot}\sim 10^{-31}$
which seems absolutely unreal. The distribution (27) gives $R_a\sim 1$. For
related problems typical suppression factor is $R\sim 1$. The process (16)
suppressed by a factor $R_b=\Gamma _b/\Gamma _{free}\approx 1/\pi $, where
the value $m_{\Phi }=\Gamma \sim 100$ MeV was used. The realistic example is a
pion production $pn\rightarrow pp\pi ^-$ in vacuum and on neutron bound in
nucleus. When $\epsilon _{\pi }$ is in the region of resonance, the pion
absorption due to interaction in the final state is very strong. This effects
on the number of pions emitted from the nucleus, but not on the fact of pion
formation inside nucleus, which is interesting for us.

(4) By means of diagram technique it is easy to get the $S$-matrix amplitude 
of potential model $M_{pot}$. From the expression for $M_{pot}$ (see Eq.(22) 
of Ref.[5]) it is seen that there is a double counting with respect to $H$. 

Points (1)-(4) suggest that the functional structure of $W_{pot}$ is wrong.
When annihilation decay probability of $\bar{n}$-nucleus $W^{\bar{n}} (\tau )=
\sum_{f\neq i}\mid T^{\bar{n}}_{fi}(\tau )\mid ^2\rightarrow 1$, the error
increases up to 100 percent resulting in change of functional structure: $W_
{ann}\sim t^2\rightarrow W_{pot}\sim t$. As $W^{\bar{n}}(\tau )\rightarrow 0$,
that is $ImU_{\bar{n}}\rightarrow 0$, the error also tends to zero and Eqs.(4)
give an exact result.

Solving Eqs.(4) by method of Green functions one can get
\begin{eqnarray}
T_{ii}(t)=i\epsilon^2 \int_{0}^{t}dt_1\int_{0}^{t_1}d\tau U^{\bar{n}}_{ii}
(\tau),\nonumber\\
T_{\bar{n}i}(t)=-\epsilon \int_{0}^{t}d\tau U^{\bar{n}}_{ii}(\tau),
\end{eqnarray}
where $U^{\bar{n}}_{ii}(\tau)$ is defined by Eq.(15). It is seen that the main
contribution gives the region $\tau <2/\Gamma $. However, in this region it is
meaningless to speak about potential $U_{\bar{n}}$. 

(The self-energy part $\Sigma=V$ is due to of zero angle rescattering of $\bar
{n}$. Such scheme is artificial, because $\sigma _{ann}^{\bar{n}N}>\sigma 
_{el}^{\bar{n}N}$ and in the first act of $\bar{n}$-medium interaction 
annihilation takes place. Also recall that for sufficiently early times the 
exponential decay law is violated [7].)

One might hope that correct result can be provided by appropriate
parametrization. However, $U_{\bar{n}}$ has been fitted to radically different 
problem. Really, the coupled Eqs.(4) give rise to the following equation:
\begin{equation}
(\partial_t^2+i\partial_t(V+2H_0)-H_0^2-H_0V+ \epsilon ^2)\Psi _n=0.
\end{equation}
$\Psi _n$ is suffice to get $W_{pot}$. Meantime, $U_{\bar{n}}$ is fitted to 
$\bar{p}$-atom and low energy scattering data. These problems are described by
stationary equations of Schrodinger type
\begin{equation}
(\nabla^2/2m+i\Gamma/2)\Psi _{\bar{n}}=-E\Psi _{\bar{n}}.
\end{equation}
The distinctions between Eqs.(29) and (30) are obvious. Eq.(29) is even not
Schrodinger type. It describes $n$ rather then $\bar{n}$. For Eq.(29) the
papameter $U_{\bar{n}}$ is uncertain; the physical sense of $ImU_{\bar{n}}$ is 
not clear. The $\Gamma $-dependence of the results is inverse: $W_{pot}\sim 1/
\Gamma $, whereas Eq.(30) gives $W\sim 1-\exp (-\Gamma t)$. The structure 
$\int_{0}^{t}d\tau U^{\bar{n}}_{ii}(\tau)$ (see Eqs.(28)) appears only in the 
oscillation problem. It is necessary to know $U^{\bar{n}}_{ii}(\tau)$ at a 
scale $\tau \sim 1/\Gamma $. So it is beyond reason to use $U_{\bar{n}}$ in 
the Eqs.(29),(4), because it was fitted to absolutely different problem.

Let us explain this point. Fitting potential $U_{\bar{n}}$ we fit the matrix 
elements associated with Eq.(30). However, these matrix elements differ
radically from those involved in Eqs.(7),(8). For example, in the Born
approximation the zero angle $\bar{p}$-nucleus scattering amplitude is $f(0)
\sim \int d^3xU_{\bar{n}}(\bf x)$. (Compare with Eqs.(28).) Thus from a pure
phenomenological standpoint $T_{ii}(t)$ is uncertain as well. It can be
calculated only by means of correct model with Hermitian operators.

For $\bar{n}$ in nuclear matter the $t$-dependent equation is
\begin{equation}
(i\partial_t-m+i\Gamma /2)\Psi _{\bar{n}}=0,
\end{equation}
$\Psi _{\bar{n}}(0)=1$. From the condition of probability conservation we have
$W_{ann}=1-\mid\!\Psi_{\bar{n}}\!\mid ^2=1-\exp (-\Gamma t)$. $\Gamma $ is
extracted from experiment. When $\Gamma $ increases, $W_{ann}$ increases as
well; so {\em it will also be for oscillation problem.} The potential model
gives inverse tendency (10). In the problem corresponding to Eq.(31) the value
of $\Gamma $ affects the branching ratio of channels only.

Taking into account above given analogy with process (3) we conclude: when 
$ImU_{\bar{n}}\neq 0$, $W_{pot}$ is wrong. The scheme based on Eqs.(12),(7) is 
improper. The parametrization $\hat{U}_{\bar{n}}\sim -i\Gamma /2$ is very 
useful for Eqs.(30),(31) i.e., for the problems with {\em prepared} $\bar{n}$.
However, it is incompatible with a realistic $\bar{n}$-nuclear dynamic 
characteristic of $n\bar{n}$ conversion.  

Finally, we point to an important difference between calculations. Since the
asymptotic of $U^{\bar{n}}_{ii}(\tau)$ is $U^{\bar{n}}_{ii}(\tau \gg 1/\Gamma )
\rightarrow 0$, the values of $T_{ii}$ and $W_{pot}$ are defined by behaviour
of $U^{\bar{n}}_{ii}(\tau)$ at a scale $\tau \sim (10^{-24}-10^{-23})$ s, which
is very "undesirably". For off-diagonal matrix elements the picture is
inverse: $T^{\bar{n}}_{f\neq i}(\tau \gg 1/\Gamma )\rightarrow const \neq 0$.
(Otherwise, the corresponding process probability is $W^{\bar{n}}_{fi}(\tau)=
\mid T^{\bar{n}}_{fi}(\tau)\mid ^2=0.$) As a result the main contribution to
Eq.(23) gives the region $1/\Gamma <\tau <t$ (rather then $\tau <2/\Gamma $),
resulting in Eqs.(26),(27). Considering off-diagonal matrix elements we obviate
the principal difficulties mentioned above.

In fact the problem is very original. From a view-point of optical potential
it is that we deal with the system of time-dependent coupled equations. In
consequence of this the results are expressed through the t-dependent matrix
elements, which should be calculated beyond the potential model. Hence we
come to the field approach. However, the oscillation Hamiltonian corresponds
to 2-tail (because of this the single-particle model is used) which leads to
infrared singularities. To avoid them the problem is considered on the finite
time interval. So we come to the time-dependent description again.

\section{Comment on Gal's paper}
In the paper [8] our result was improperly "corrected". The probability to 
find the annihilation products is $W_{ann}(t)\approx \epsilon^2t^2W^{\bar{n}}$
(see Eq.(27)), where $W^{\bar{n}}$ is the $\bar{n}$-nucleus decay probability
$W^{\bar{n}}=1-\exp(-\Gamma t)\approx 1$; $\epsilon^2t^2$ is the free-space
$n\bar{n}$ transition probability. "Correction" is that instead of $W^{\bar{n}
}$ the probability to find $\bar{n}$ (Eq.(6)) is substituted. As a result $W_
{ann}(t)\approx \epsilon^2t^2W_{\bar{n}}$. (See Eqs.(31)-(33) of Ref.[8].) We
read: The probability to find annihilation products=(the $n\bar{n}$ transition
probability)$\times $(the probability to find $\bar{n}$-nucleus). Obviously,
this "correction" is only trivial error. The rest of the "results" [8] are
known since 1980. For example, Eq.(27) of Ref.[8] (the final result) coincides
with Eq.(2) of Ref.[5] (the beginning of this paper). In the Ref.[8] the Eqs.
(4) are solved. This solution is known since Eiler's time. The essence of the
problem is in the description of the dynamic of the process, i.e., in
approximation (12).

In the abstract [8] we read:"...within a simple model which respects 
unitarity". Unitarity within a model with "anti-Hermitian" operator $U_{\bar
{n}}=-i\Gamma /2$? 

Also Gal writes: "Dover, Gal, Richard [9] pinpointed errors in Ref.[4]." In 
fact, the situation is inverse. The simple calculation given in [9] confirms 
the FTA and was already done in our first paper [4] (zero angle rescattering 
diagrams of $\bar{n}$) as well as in Ref.[5] (see Pgs. R1884, R1885). Since 
the authors [8,9] "overlooked" all of this we cite here only one paragraph 
from Ref.[5]: "The authors [9] substitute $H=-i\Gamma /2$ in Eq.(4) of Ref.[5]
and obtain $W_{pot}$. On the basis of this and only this they refute the 
result of Ref.[4]. In other words they refute our result because it differs
from the potential model one."

Unusual logic. The procedure mentioned above is a verification of FTA by the 
example of exactly solveble (but incorrect) potential model. We abandon the 
approximation (12), i.e., the potential description of $\bar{n}$-medium
interaction in principal, what was clearly pointed in [4,5].

\section{On an anhancement of oscillations}
In the region $\Gamma t\gg 1$
$$
W_{ann}(t)/W_{pot}(t)\sim -ImVt\sim \Gamma t\gg 1
$$
what means an anhancement of oscillations as compared with standard result.
Certainly, this is a record value. For other problems an enhancement factor
will provoke a lesser skepticism. In fact the suppression factor
$R_{pot}\sim 10^{-31}$ is surprising.

When $\Gamma t<1$, annihilation products as well as $\bar{n}$ in the final
state can be detected. In this case condition (24) is not fulfilled.
Nevertheless, the qualitative picture remains as before: in the standard
approach $ImU_{\bar{n}}$ leads to suppression of oscillations, in our one this
is not the case. The presence of open channels of ${\bar{n}}$-medium
interaction (which are described by $ImU_{\bar{n}}$ in the potential model)
do not lead to suppression  of oscillations.

For large part of actual problems the both approaches give an identical result.
For two-step processes of the type (1), when $b$-particle absorption is
essential, the principal disagreement takes place. We do not want to make the
categorical conclusions. Nevertheless, two points are obvious: (1) The
potential approach should be considered only as starting point. (2) The only
assumption used in our calculation is expression (24). However, this condition
is basic to the all $S$-matrix calculations.  

\bigskip
{\large\bf Figure caption}
\\
Fig.1. (a) $n\bar{n}$ transition in a medium followed by annihilation;
(b) The same as in Fig.(a), but with escaping of particle in the $n\bar{n}$
transition vertex.
\\
Fig.2. Free-space process $n\rightarrow \bar{n}\rightarrow \bar{p}e^+\nu$.

\bigskip
\begin{thebibliography}{99}
\bibitem{1}
M.L. Good, Phys. Rev. {\bf 106} (1957) 591; 

Phys. Rev. {\bf 110} (1958) 550;

L. Wolfenstein, Phys. Rev. {\bf D17} (1978) 2369; 

E.D. Commins and P.H. Bucksbaum,{\em Weak Interactions of Leptons and Quarks} 
(Cambridge University Press, 1983); 

F. Boehm and P. Vogel, {\em Physics of Massive Neutrinos} (Cambridge
University Press, 1987).
%
\bibitem{2}
V.A. Kuzmin, JETF Lett.{\bf 12} (1970) 228;

S.L. Glashow, Preprint HUTP-79/A059 (Harvard, 1979);

R.E. Marshak and R.N. Mohapatra, Phys. Rev. Lett. {\bf 44} (1980) 1316. 
%
\bibitem{3}
M.V. Kazarnovsky, V.A. Kuzmin and M.E. Shaposhnikov, Pis'ma Zh.Exsp.Theor.Fiz.
{\bf 32} (1980) 88; 

K.G. Chetyrkin, M.V. Kazarnovsky, V.A. Kuzmin and M.E. Shaposhnikov,
Phys. Lett. {\bf B99} (1981) 358;

P.G.H. Sandars, J.Phys. {\bf G6} (1980) L161;

V.I. Nazaruk, Yad.Fiz. {\bf 56} (1993) 153.
%
\bibitem{4}
V.A. Kuzmin and V.I. Nazaruk, in: Proc. of the III International
Symp. on Weak and Electromagnatic Inter. in Nuclei, Dubna 1992, p.449.

V.I. Nazaruk, Phys. Lett. {\bf B337} (1994) 328.
%
\bibitem{5}
V.I. Nazaruk, Phys. Rev. {\bf C58} (1998) R1884;

V.I. Nazaruk, in: Proc. of the 7th Conference Intersections of Particle
and Nuclei Physics, CIPANP2000, Quebec City, Canada, May 2000, p.901.
%
\bibitem{6}
H. Takita et al.(Kamiokande), Phys.Rev.{\bf D34} (1986) 902.
%
\bibitem{7}
L.A. Khalfin, Phys.Lett. {\bf B112} (1982) 223; 

G.N. Fleming, Phys.Lett.{\bf B125} (1983) 187; 

K. Grotz and H.V. Klapdor, Phys.Rev. {\bf C30} (1984) 2098.
%
\bibitem{8}
A. Gal, Phys. Rev. {\bf C61} (2000) 028201.
%
\bibitem{9}
C.B.Dover, A.Gal and J.M.Richard, Phys. Lett. B {\bf 344} (1995) 433.
%
\end{thebibliography}
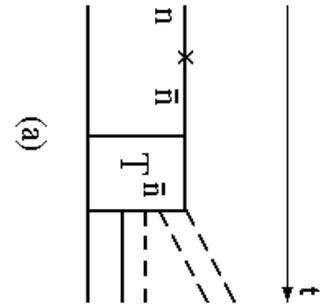
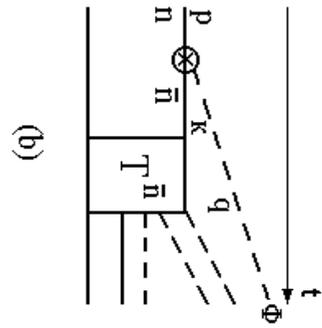
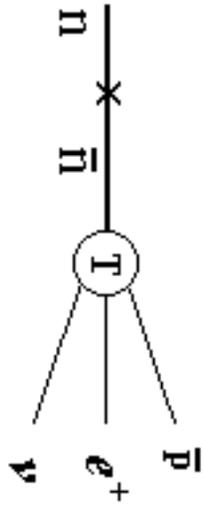
\end{document}